\documentclass{eurodyn2011}
\usepackage[pdftex]{graphicx}
\usepackage{amsmath}
\usepackage{amssymb}

\title{A damage-plasticity model for the dynamic failure of concrete}

\author{ P. Grassl$^1$, U. Nystr\"{o}m$^2$, R. Rempling$^2$, K. Gylltoft$^2$\\
$^1$School of Engineering, University of Glasgow, Glasgow G128LT, Scotland, UK\\
$^2$Department of Civil and Environmental Engineering, Chalmers University of Technology, S-41296 G\"{o}teborg, Sweden\\
email: peter.grassl@glasgow.ac.uk,\{ulrika.nystrom, rasmus.rempling, kent.gylltoft\}@chalmers.se \\ \vspace{0.5cm}
Preprint. Submitted to Eurodyn 2011, 8th International Conference on Structural Dynamics, Leuven, Belgium, 2011}

\abstract{
A constitutive model based on the combination of  damage mechanics and plasticity is developed to analyse concrete structures subjected to dynamic loading.
The aim is to obtain a model, which requires input parameters with clear physical meanings.
The model should describe the important characteristics of concrete subjected to multiaxial and rate-depending loading. 
This is achieved by combining an effective stress based plasticity model with an isotropic damage model based on plastic and elastic strain measures.
The model response in tension, uni-, bi- and tri-axial compression is compared to experimental results in the literature.}


\begin{document}

\maketitle

\section{Introduction}

Concrete is a strongly heterogeneous material, which exhibits a complex nonlinear mechanical behaviour.
Failure in tension and low confined compression is characterised by softening which is defined as decreasing stress with increasing deformations.
This softening response is accompanied by a reduction of the unloading stiffness of concrete, and irreversible (permanent) deformations, which are localised in narrow zones often called cracks or shear bands.
On the other hand, the behaviour of concrete subjected to high confined compression is characterised by a ductile hardening response; that is, increasing stress with increasing deformations.
Furthermore, high loading rates are known to significantly increase the strength in tension and compression.
These phenomena should be considered in a constitutive model for analysing the dynamic behaviour of concrete structures.

There are many constitutive models for the nonlinear response of concrete proposed in the literature. 
Commonly used frameworks are plasticity, damage mechanics and combinations of plasticity and damage mechanics.
Stress-based plasticity models are useful for the modelling of concrete subjected to triaxial stress states, since the yield surface corresponds at a certain stage of hardening to the strength envelope of concrete. 
Furthermore, the strain split into elastic and plastic parts represents realistically the observed deformations in confined compression, so that unloading and path-dependency can be described well.
However, plasticity models are not able to describe the reduction of the unloading stiffness that is observed in experiments.
Conversely, strain based damage mechanics models are based on the concept of a gradual reduction of the elastic stiffness driven by strain measures. For isotropic damage mechanics models, the stress evaluation procedure is explicit, which allows for a direct determination of the stress state, without an iterative calculation procedure. Furthermore, the stiffness degradation in tensile and low confined compressive loading observed in experiments can be described.
However, isotropic damage mechanics models are often unable to describe irreversible deformations observed in experiments and are mainly limited to tensile and low confined compression stress states \cite{Mazars84}.
On the other hand, combinations of isotropic damage and plasticity are widely used and many different models have been proposed in the literature.
One popular class of models relies on a combination of stress-based plasticity formulated in the effective stress space combined with a strain based damage model \cite{Ju89,LeeFen98,JasHuePijGha06,GraJir06,GraRem08,Gra09b}.

In the present work, the combined damage-plasticity model presented in \cite{GraJir06,GraJir06a} (CDPM1) is revisited to develop an constitutive model for the rate dependent failure of concrete, which is characterised by its numerical stability, well defined input parameters and flexibility to be adapted to newly developed concrete based materials, such as fibre reinforced concrete.
As this model can be seen as an augmentation of CDPM1 it is called here CDPM2.
The stress-based plasticity part of the model is based on the effective stress.
It includes hardening in the post-peak regime, which is used to model the strain rate dependence of strength by delaying the onset of damage.
The plasticity part is combined with a damage model, which is based on elastic and plastic strain measures and distinguishes between tensile and compressive stress states using an approach similar to those proposed in \cite{Mazars84,Ort87,FicBorPij99}. The damage model is used to describe the complex strength envelope of concrete. With this combination of plasticity and damage mechanics, it is aimed to provide a computationally efficient model for the dynamic behaviour of concrete.

\section{Model}

\subsection{General framework} \label{sec:General}

The damage plasticity constitutive model is based on the following stress-strain relationship:
\begin{equation}\label{eq:general}
\boldsymbol{\sigma} = \left(1-\omega_{\rm t}\right) \bar{\boldsymbol{\sigma}}_{\rm t} + \left(1-\omega_{\rm c}\right) \bar{\boldsymbol{\sigma}}_{\rm c}
\end{equation}
where $\bar{\boldsymbol{\sigma}}_{\rm t}$ and $\bar{\boldsymbol{\sigma}}_{\rm c}$ are the positive and negative parts of the effective stress tensor $\bar{\boldsymbol{\sigma}}$, respectively, and $\omega_{\rm t}$ and $\omega_{\rm c}$ are two scalar damage parameters, ranging form 0 (undamaged) to 1 (fully damaged).
The effective stress $\bar{\boldsymbol{\sigma}}$ is defined as
\begin{equation}
\bar{\boldsymbol{\sigma}} = \mathbf{D}_{\rm e} : \left(\boldsymbol{\varepsilon} - \boldsymbol{\varepsilon}_{\rm p}\right) 
\end{equation}
where $\mathbf{D}_{\rm e}$ is the elastic stiffness tensor based on the elastic Young's modulus $E$ and Poisson's ratio $\nu$, $\boldsymbol{\varepsilon}$ is the strain tensor and $\boldsymbol{\varepsilon}_{\rm p}$ is the plastic strain tensor. The positive and negative parts of the effective stress $\bar{\boldsymbol{\sigma}}$ in Eq.~(\ref{eq:general}) are determined from the principal effective stress as $\bar{\boldsymbol{\sigma}}_{\rm pt} = \left< \bar{\boldsymbol{\sigma}}_{\rm p} \right>_+$ and $\bar{\boldsymbol{\sigma}}_{\rm pc} = \left< \bar{\boldsymbol{\sigma}}_{\rm p} \right>_-$, where $\left< \right>_+$ and $\left< \right>_-$ are positive and negative part operators, respectively, defined as $\left< x\right>_+ = \max\left(0,x\right)$ and $\left< x\right>_- = \min\left(0,x\right)$.
In addition, a scalar measure $\alpha_{\rm c}$ is defined which distinguishes between tensile and compressive stress states.
It has the form
\begin{equation} \label{eq:alpha}
\alpha_{\rm c} = \sum_i \dfrac{\left<\bar{\sigma}_{{\rm p}i}\right>_-\left( \left<\bar{\sigma}_{{\rm p}i}\right>_+ + \left<\bar{\sigma}_{{\rm p}i}\right>_- \right)}{\| \bar{\boldsymbol{\sigma}}_{\rm p} \|^2}
\end{equation}
where $\left<\bar{\sigma}_{\rm pi}\right>_+$ and $\left<\bar{\sigma}_{\rm pi}\right>_-$ are the components of the compressive and tensile part of the principal stresses, respectively.
The parameter $\alpha_{\rm c}$ varies in the range from 0 to 1. For instance, for a combined tensile and compressive stress state with principal stress components $\bar{\boldsymbol{\sigma}}_{\rm p} = \left( - \bar{\sigma}, 0.2 \bar{\sigma}, 0.1 \bar{\sigma} \right)^{\rm T}$, the positive and negative principal stresses are $\boldsymbol{\bar{\sigma}}_{\rm pt} = \left( 0, 0.2 \bar{\sigma}, 0.1 \bar{\sigma} \right)^{\rm T}$ and $\boldsymbol{\bar{\sigma}}_{\rm pc} = \left( -\bar{\sigma}, 0, 0 \right)^{\rm T}$, respectively. For this stress state, the variable is $\alpha_{c} = 0.95$.
This measure is used later in the definition of the damage parameter.

The plasticity model is based on the effective stress and thus independent of damage. 
The model is described by the yield function, the flow rule, the evolution law for the hardening variable and the loading-unloading conditions. 
The form of the yield function is
\begin{equation} \label{eq:yield}
f_{\rm p} \left(\bar{\boldsymbol{\sigma}}, \kappa_{\rm p} \right) = F \left(\bar{\boldsymbol{\sigma}}, q_{\rm h}, q_{\rm s}\right)
\end{equation} 
where $q_{\rm h}$ and $q_{\rm s}$  are hardening functions, which depend on the plastic hardening parameter $\kappa_{\rm p}$.
The flow rule is
\begin{equation}\label{eq:flowRule}
\dot{\boldsymbol{\varepsilon}}_{\rm p} = \dot{\lambda} \dfrac{\partial g_{\rm p}}{\partial \bar{\boldsymbol{\sigma}}}
\end{equation}
where $\dot{\boldsymbol{\varepsilon}}_{\rm p}$ is the rate of the plastic strain, $\dot{\lambda}$ is the rate of the plastic multiplier and $g_{\rm p}$ is the plastic potential.
The rate of the hardening parameter $\kappa_{\rm p}$ is related to the rate of the plastic strain by an evolution law, which is presented in Section~\ref{sec:plast}.
The loading-unloading conditions are
\begin{equation}\label{eq:UnloadPlast}
f_{\rm p}\leq 0, \hskip 5mm \dot{\lambda} \geq 0, \hskip 5mm \dot{\lambda} f_{\rm p} = 0
\end{equation}
A detailed description of the individual components of the plasticity model are discussed in Section~\ref{sec:plast}

The damage part of the present damage-plasticity model is related to elastic and plastic strain measures.
For a pure tensile stress state, $\bar{\boldsymbol{\sigma}}_{\rm t} = \bar{\boldsymbol{\sigma}}$, $\bar{\boldsymbol{\sigma}}_{\rm c}$ is zero and $\omega_{\rm t} = \omega$, so that Eq.~(\ref{eq:general}) becomes 
\begin{equation}
\boldsymbol{\sigma} = \left(1-\omega\right) \bar{\boldsymbol{\sigma}} = \left(1-\omega\right) \mathbf{D}_{\rm e} : \left( \boldsymbol{\varepsilon} - \boldsymbol{\varepsilon}_{\rm p} \right)
\end{equation}

\begin{figure}
\begin{center}
\includegraphics[width=0.7\linewidth]{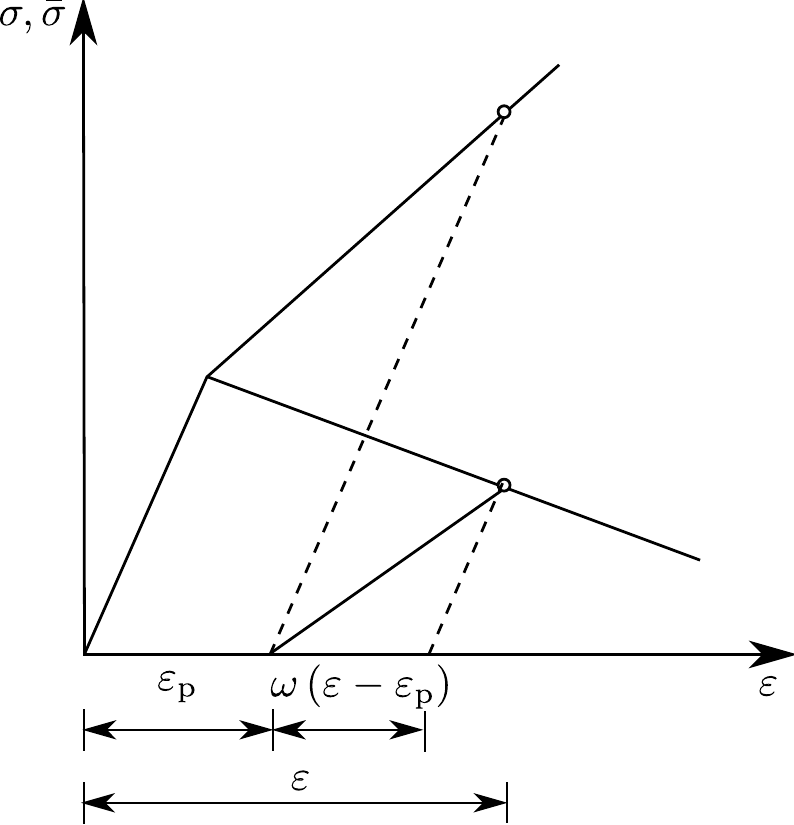}
\end{center}
\caption{Geometrical meaning of the inelastic strain \protect $\varepsilon_{\rm i}$ for the combined damage-plasticity model. The inelastic strain is composed of reversible \protect $\omega \left(\varepsilon - \varepsilon_{\rm p}\right)$ and irreversible $\varepsilon_{\rm p}$ parts. The dashed lines represent elastic unloading with the same stiffness as the initial elastic loading.}
\label{fig:compExplain}
\end{figure}

The equation is rearranged as 
\begin{equation}
\boldsymbol{\sigma} = \mathbf{D}_{\rm e} : \left(\boldsymbol{\varepsilon} - \left(\boldsymbol{\varepsilon}_{\rm p} + \omega \left(\boldsymbol{\varepsilon}-\boldsymbol{\varepsilon}_{\rm p}\right)\right)\right) = \mathbf{D}_{\rm e} : \left(\boldsymbol{\varepsilon} - \boldsymbol{\varepsilon}_{\rm i}\right)
\end{equation}
where
\begin{equation}\label{eq:inelastic}
\boldsymbol{\varepsilon}_{\rm i} = \boldsymbol{\varepsilon}_{\rm p} + \omega \left( \boldsymbol{\varepsilon} - \boldsymbol{\varepsilon}_{\rm p} \right)
\end{equation}
is the inelastic strain which is subtracted from the elastic strain.
The geometrical interpretation of the inelastic strain and its split for uniaxial tension, hardening plasticity and linear damage evolution are shown in Fig.~\ref{fig:compExplain}. The part $\omega \left(\boldsymbol{\varepsilon} - \boldsymbol{\varepsilon}_{\rm p}\right)$ is reversible and $\boldsymbol{\varepsilon}_{\rm p}$ is irreversible.
These two parts of the inelastic strain are used to define the damage history variables, see Section~\ref{sec:dam}.

\subsection{Plasticity part} \label{sec:plast}
The plasticity model is formulated in a three-dimensional framework with a pressure-sensitive yield surface, hardening and non-associated flow. 
The main components are the yield condition, the hardening law, the evolution law for the hardening variable and the flow rule. 
This model is an extension of CDPM1 in \cite{GraJir06}, in which the plasticity response is assumed to be perfect-plastic in the regime in which damage is active. 
In the present study, the plasticity part exhibits hardening in this regime, which is used to model the strain rate dependence of the strength by adjusting the onset of damage.
This extension to the hardening requires several extensions of the plasticity part of CDPM1, which is presented in the following section.

The yield surface is described in terms of the cylindrical coordinates in the principal effective stress space (Haigh-Westergaard coordinates), which are the volumetric effective stress $\bar{\sigma}_{\rm V}$, the norm of the deviatoric stress $\bar{\rho}$ and the Lode angle of the deviatoric stress $\bar{\theta}$. For a definition of these coordinates it is referred to \cite{GraJir06}.

The yield function
\begin{equation} \label{eq:yieldSurface}    
\begin{split}
& f_{\rm p}(\bar{\sigma}_{\rm V},\bar\rho,\bar\theta;\kappa_{\rm p})= \\
& \left\{\left[1-q_{\rm{h}}(\kappa_{\rm p})\right]\left( \frac{\bar{\rho}} {\sqrt{6}f_{\rm c}} + \frac{\bar{\sigma}_{\rm V}} {f_{\rm c}} \right)^2 + \sqrt{\frac{3}{2}} \frac {\bar{\rho}}{f_{\rm c}} \right\}^2 \\
& +m_0 q_{\rm{h}}(\kappa_{\rm p})q_{\rm{s}}(\kappa_{\rm p}) \left[\frac{\bar{\rho} }{\sqrt{6}f_{\rm c}}r(\cos{\bar{\theta}}) + \frac{\bar{\sigma}_{\rm V}}{f_{\rm c}} \right]\\
& - q_{\rm{h}}^2(\kappa_{\rm p}) q_{{\rm s}}^2(\kappa_{\rm p})
\end{split}
\end{equation}
depends on the effective stress (which enters in the form of cylindrical coordinates) and on the hardening variable $\kappa_{\rm p}$ (which enters through the dimensionless variables $q_{\rm h}$ and $q_{\rm s}$). Parameter $f_{\rm c}$ is the uniaxial compressive strength. 

\begin{figure}
\begin{center}
\includegraphics[width=0.7\linewidth]{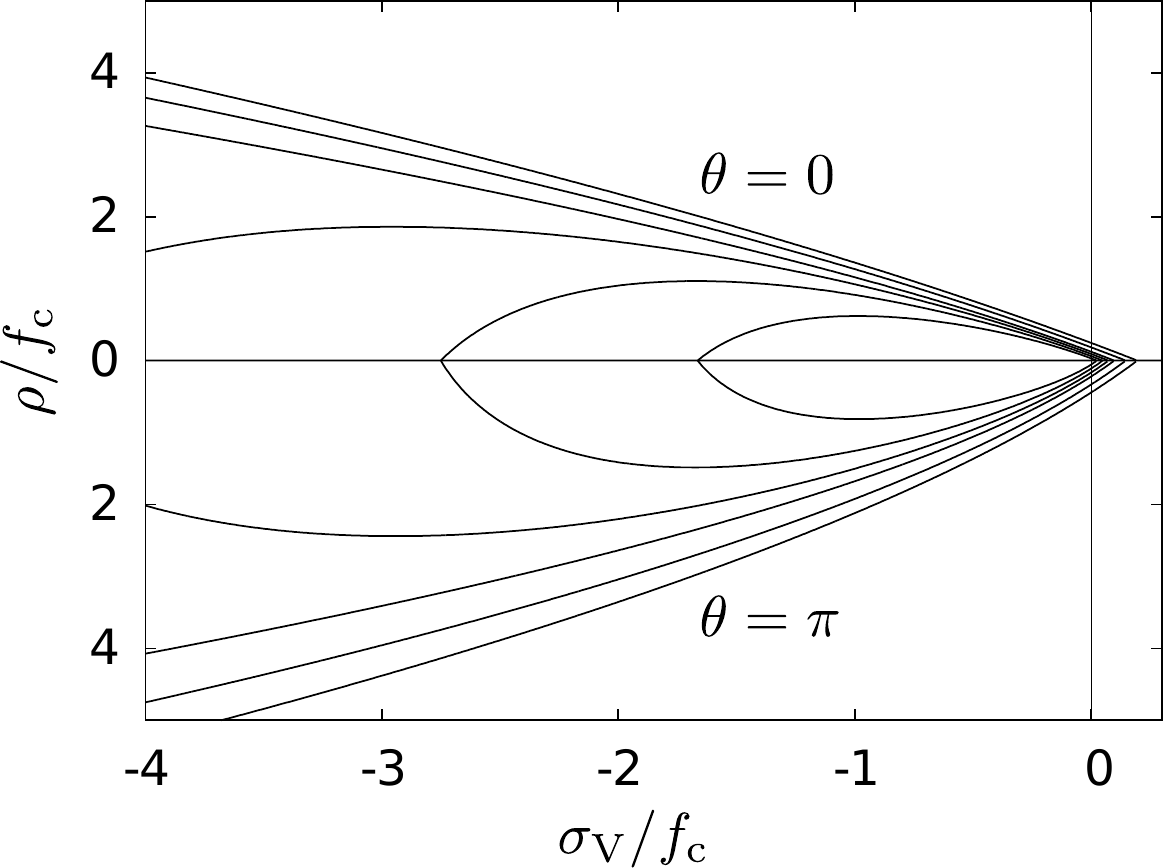}
\end{center}
\caption{The evolution of the meridional section of the yield surface during hardening in the pre-peak and post-peak regime.}
\label{fig:surfaceMeridian}
\end{figure}

\begin{figure}
\begin{center}
\includegraphics[width=0.7\linewidth]{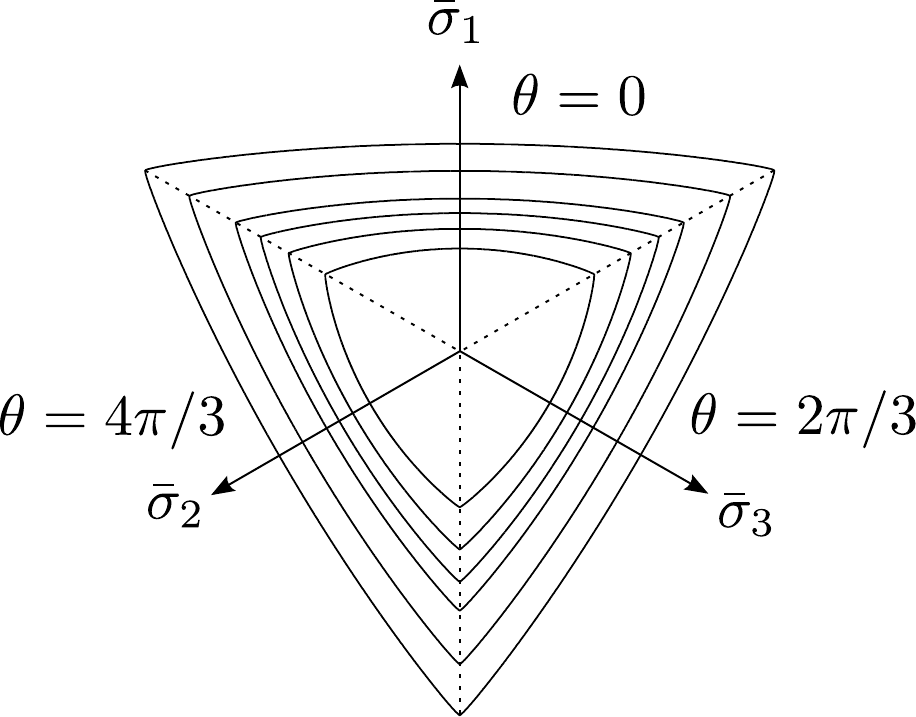}
\end{center}
\caption{The evolution of the deviatoric section of the yield surface during hardening for a constant volumetric stress of $\bar{\sigma}_{\rm V} = - f_{\rm c}/3$.}
\label{fig:surfaceDeviatoric}
\end{figure}

\begin{figure}
\begin{center}
\includegraphics[width=0.7\linewidth]{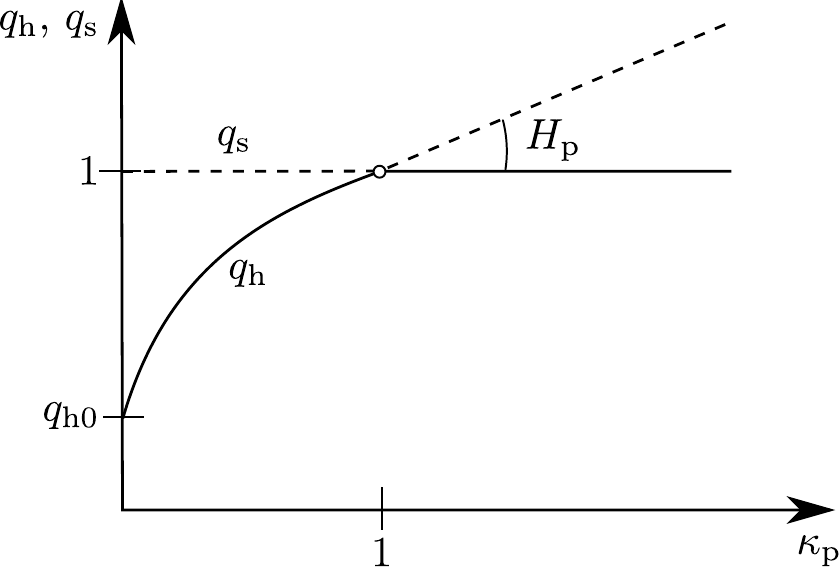}
\end{center}
\caption{The two hardening laws $q_{\rm h}$ (solid line) and $q_{\rm s}$ (dashed line).}
\label{fig:hardening}
\end{figure}

The meridians of the yield surface $f_{\rm p}=0$ are parabolic, and the deviatoric sections change from triangular shapes at low confinement to almost circular shapes at high confinement.
The shape of the deviatoric section is controlled by the function
\begin{equation} \label{eq:rFunction}
\begin{split}
&r(\cos{\bar{\theta}}) =\\
& \frac{4(1-e^2)\cos^2{\bar{\theta}} + (2e-1)^2}{2(1-e^2)\cos{\bar{\theta}} + (2e-1)\sqrt{4(1-e^2)\cos^2{\bar{\theta}}+5e^2 -4e}} 
\end{split}
\end{equation} 
proposed by \cite{Willam74}. 
The eccentricity parameter $e$ and the friction parameter 
\begin{equation}\label{eq:frictionM}
m_0 = \dfrac{3 \left(f_{\rm c}^2 - f_{\rm t}^2\right)}{f_{\rm c}f_{\rm t}} \dfrac{e}{e+1}
\end{equation}
are calibrated from the values of uniaxial and equibiaxial compressive strengths and uniaxial tensile strength as described in \cite{JirBaz01}.
The shape of the meridians of the yield surface is controlled by the hardening variables $q_{\rm h}$ and $q_{\rm s}$ and $m_0$.
The evolution of the yield surface during hardening is presented in Figure~\ref{fig:surfaceMeridian}~and~\ref{fig:surfaceDeviatoric}.
The two hardening functions, which are functions of the hardening variable $\kappa_{\rm p}$ are presented in Figure~\ref{fig:hardening}. 
If the two variables $q_{\rm h}$ and $q_{\rm s}$ are equal to one, the yield surface turns into the failure surface proposed by~\cite{Menetrey95}.

In the present model, the flow rule in Eq.~(\ref{eq:flowRule}) is non-associated, which means that the yield function $f_{\rm p}$ and the plastic potential $g_{\rm p}$ do not coincide and, therefore, the direction of the plastic flow $\mathbf{m}\equiv\partial g_{\rm p}/\partial\bar{\boldsymbol{\sigma}}$ is not normal to the yield surface.
This is important for realistic modeling of the volumetric expansion under compression for frictional materials such as concrete. An associated flow rule for this type of yield surface gives an unrealistically high volumetric expansion in compression, which leads in the case of passive confinement to an overestimated strength (peak stress); see \cite{Gra04}. 
The plastic potential is given as
\begin{equation} \label{eq:plasticPotential}
\begin{split}
& g_{\rm p}(\bar\sigma_{\rm V},\bar\rho;\kappa_{\rm p})=\\
& \left\{\left[1-q_{\rm{h}}(\kappa_{\rm p})\right] \left( \frac{\bar{\rho}} {\sqrt{6}f_{\rm c}} + \frac{\bar{\sigma}_{\rm V}}{f_{\rm c}} \right)^2 + \sqrt{\frac{3}{2}} \frac {\bar{\rho}}{f_{\rm c}} \right\}^2\\
& + q_{\rm{h}}(\kappa_{\rm p}) q_{\rm{s}}(\kappa_{\rm p}) \left( \frac{m_0 \bar{\rho}}{\sqrt{6}f_{\rm c}} + \frac{m_{\rm g}(\bar{\sigma}_{\rm V}, \kappa_{\rm p})}{f_{\rm c}} \right)
\end{split}
\end{equation}
where
\begin{equation} \label{eq:mg}
m_{\rm g}(\bar{\sigma}_{\rm V}, \kappa_{\rm p})=A_{\rm g}\left(\kappa_{\rm p}\right)B_{\rm g}\left(\kappa_{\rm p}\right)f_{\rm c} \exp{\frac{\bar{\sigma}_{\rm V} - q_{\rm{s}}(\kappa_{\rm p}) f_{\rm t}/3}{B_{\rm g}\left(\kappa_{\rm p}\right) f_{\rm c}}}
\end{equation}
is a variable controlling the ratio of volumetric and deviatoric plastic flow.
The variables $A_{\rm g}\left(\kappa_{\rm p}\right)$ and $B_{\rm g}\left(\kappa_{\rm p}\right)$, which depend on the hardening function $q_{\rm s}$, are derived from assumptions on the plastic flow in uniaxial tension and compression.
In uniaxial tension, the laterial plastic strain is chosen to be zero. In uniaxial compression, the ratio of axial and laterial plastic strain is set to the model constant $D_{\rm f}$.
The plastic potential does not depend on the third Haigh-Westergaard coordinate (Lode angle $\bar{\theta}$). 
This increases the efficiency of the implementation and the robustness of the model. 
However, it also limits the capability of this flow rule to describe the response of concrete in multiaxial compression.

The evolution law for the hardening variable $\kappa_{\rm p}$ sets the rate of the hardening variable equal to the norm of the plastic strain rate scaled by a hardening ductility measure.
This scaling factor is constructed such that the model response is more ductile under compression \cite{GraJir06}. 
 
\subsection{Damage part} \label{sec:dam}
The damage parameter is determined by means of history variables that are based on measures of the plastic and elastic strain.
The measure of the plastic strain is based on the norm of the rate of the plastic strain.
The measure of the elastic strain, denoted here as the equivalent strain $\tilde{\varepsilon}$, is more difficult to choose. 
With the equivalent strain, the onset of damage is determined.
Thus, it is required to choose an equivalent strain measure, which represents at a certain value stress states which are located on the strength envelope of concrete. 
In the present work, the equivalent strain $\tilde{\varepsilon}$ is chosen so that the effective stress states at the onset of damage satisfy the strength criterion developed by Menetrey and Willam \cite{Menetrey95} on which the yield surface in the plasticity part is based.
The corresponding damage envelope has the form 
\begin{equation}\label{eq:damageStrength}  
\dfrac{3}{2} \dfrac{\bar{\rho}^2}{f_{\rm c}^2} + q_{\rm d} m_0 \left(\frac{\bar{\rho} }{\sqrt{6}f_{\rm c}}r\left(\cos \bar{\theta}\right) + \dfrac{\bar{\sigma}_{\rm V}}{f_{\rm c}} \right) - q_{\rm d}^2 = 0
\end{equation}
From Eq.~(\ref{eq:damageStrength}), the variable $q_{\rm d}$ (positive root) is determined, which is used to define the equivalent strain $\tilde{\varepsilon}$ for the strain driven damage model.
The meaning of $q_{\rm d}$ is illustrated by two representative effective stress states, namely uniaxial tension and uniaxial compression.
For uniaxial tension, the effective stress state is defined as $\bar{\sigma}_1=\bar{\sigma}_{\rm t}$, $\bar{\sigma}_2=\bar{\sigma}_3=0$, $\bar{\sigma}_{\rm V} = \bar{\bar{\sigma}}_{\rm t}/3$, $\bar{s}_1 = 2 \bar{\sigma}_{\rm t}/3$, $\bar{s}_2 = \bar{s}_2 = - \bar{\bar{\sigma}}_{\rm t}/3$ and $\bar{\rho} = \sqrt{2/3} \bar{\sigma}_{\rm t}$ and $\cos \bar{\theta} = 1$.
 Setting this in Eq.~(\ref{eq:damageStrength}) results in
\begin{equation}
f_{\rm d}\left(\bar{\sigma}_{\rm t}\right) = \dfrac{\bar{\sigma}_{\rm t}^2}{f_{\rm c}^2} + q_{\rm s} m_0 \left(\dfrac{r\left(\cos \bar{\theta}\right)}{3 f_{\rm c}} + \dfrac{1}{3 f_{\rm c}}\right) \bar{\sigma}_{\rm t} - q_{\rm s}^2 = 0
\end{equation}
With $r = \dfrac{1}{e}$ for $\theta=0$ and the definition of $m_0$ in Eq.~(\ref{eq:frictionM}), this simplifies to
\begin{equation}
f_{\rm d} \left(\bar{\sigma}_{\rm t}\right) = \dfrac{\bar{\sigma}_{\rm t}^2}{f_{\rm c}^2} + q_{\rm d} \bar{\sigma}_{\rm t} \dfrac{f_{\rm c}^2-f_{\rm t}^2}{f_{\rm c}^2f_{\rm t}} - q_{\rm d}^2 = 0
\end{equation}
The positive root of this equation for $q_{\rm d}$ is
\begin{equation}
q_{\rm d} = \dfrac{\bar{\sigma}_{\rm t}}{f_{\rm t}}
\end{equation}

On the other hand, for uniaxial compression, the effective stress state is defined as $\bar{\sigma}_1 = -\bar{\sigma}_{\rm c}$, $\bar{\sigma}_2 = \bar{\sigma}_3 = 0$, $\bar{\sigma}_{\rm V} = -\bar{\sigma}_{\rm c}/3$ and $\bar{\rho} = \sqrt{\dfrac{2}{3}} \bar{\sigma}_{\rm c}$. 
Repeating the steps, which were presented above for uniaxial tension, the damage hardening parameter is determined as
\begin{equation}
q_{\rm d} = \dfrac{\bar{\sigma}_{\rm c}}{f_{\rm c}}
\end{equation} 
Consequently, $q_{\rm d}$ is proportional to the effective stresses in tension and compression, respectively, and is equal to one, if the effective stresses are equal to the strength described by the criterion in Eq.~(\ref{eq:damageStrength}). 
Therefore, the variable $q_{\rm d}$ is well suited to be used for the definition of the equivalent strain, which is
\begin{equation}
\tilde{\varepsilon} =  \dfrac{f_{\rm t}}{E} q_{\rm d}
\end{equation}   

Based on the plastic hardening parameter $\kappa_{\rm p}$, and the equivalent strain $\tilde{\varepsilon}$, six damage history variables ($\kappa_{\rm dt}$, $\kappa_{\rm dc}$, $\kappa_{\rm d1t}$, $\kappa_{\rm d1c}$,  $\kappa_{\rm d2t}$, $\kappa_{\rm d2c}$) are defined.
The evolution of these damage variables is controlled by two damage loading functions
\begin{equation} \label{eq:tensileLoading}
f_{\rm dt} = \dfrac{1}{\alpha_{\rm r}} \tilde{\varepsilon} - \kappa_{\rm dt}
\end{equation}
and 
\begin{equation} \label{eq:compLoading}
f_{\rm dc} = \dfrac{\alpha_{\rm c}}{\alpha_{\rm r}} \tilde{\varepsilon} - \kappa_{\rm dc}
\end{equation}
for which the loading-unloading conditions
\begin{equation} \label{eq:kappadt}
f_{\rm dt}\leq 0 \mbox{,} \hspace{0.5cm} \dot{\kappa}_{\rm dt} \geq 0 \mbox{,} \hspace{0.5cm} \dot{\kappa}_{\rm dt} f_{\rm dt} = 0 
\end{equation}
and
\begin{equation}\label{eq:kappadc}
f_{\rm dc}\leq 0 \mbox{,} \hspace{0.5cm} \dot{\kappa}_{\rm dc} \geq 0 \mbox{,} \hspace{0.5cm} \dot{\kappa}_{\rm dc} f_{\rm dc} = 0 
\end{equation}
apply, respectively.
With the variable $\alpha_{\rm r}$ in Eqs.~(\ref{eq:tensileLoading})~and~(\ref{eq:compLoading}) the strain rate dependence of the material is modelled, see Sec.~\ref{sec:rate}.
Furthermore, $\alpha_{\rm c}$ from Eq.~(\ref{eq:alpha}) is used to distinguish between tensile and compressive stress states.
The next two variables are linked to the plastic hardening parameter. Here, only plastic strains after the onset of damage are considered. 
They are defined as
\begin{equation}\label{eq:kappad1t}
\dot{\kappa}_{\rm d1t} = \left \{ \begin{array}{ll}
\dfrac{1}{x_{\rm s} \alpha_{\rm r}} \dot{\kappa}_{\rm p} & \mbox{if $\dot{\kappa}_{\rm dt} > 0$ $\land$ $\kappa_{\rm dt} >\varepsilon_0$}\\
0 & \mbox{if $\dot{\kappa}_{\rm dt} = 0$ $\lor$ $\kappa_{\rm dt} <\varepsilon_0$}
\end{array} \right.
\end{equation}
and
\begin{equation}\label{eq:kappad1c}
\dot{\kappa}_{\rm d1c} = \left \{ \begin{array}{ll}
\dfrac{\alpha_{\rm c}}{x_{\rm s} \alpha_{\rm r}} \dot{\kappa}_{\rm p} & \mbox{if $\dot{\kappa}_{\rm dc} > 0$ $\land$ $\kappa_{\rm dc} >\varepsilon_0$}\\
0 & \mbox{if $\dot{\kappa}_{\rm dc} = 0$ $\lor$ $\kappa_{\rm dc} <\varepsilon_0$}
\end{array} \right.
\end{equation}
where $x_{\rm s}$ is a ductility measure, which describes the influence of multiaxial stress states on the softening response and controls so the ratio of fracture energies in tension  $G_{\rm ft}$ and compression $G_{\rm fc}$.
These two history variables represent the irreversible component of the inelastic strain introduced in Sec.~\ref{sec:General}.
Finally, the last two damage history variables are related to the maximum equivalent strains $\kappa_{\rm dt}$ and $\kappa_{\rm dc}$ and are defined as
\begin{equation}\label{eq:kappad2t}
\dot{\kappa}_{\rm d2t} = \dfrac{\dot{\kappa}_{\rm dt}}{x_{\rm s}}
\end{equation}
and
\begin{equation}\label{eq:kappad2c}
\dot{\kappa}_{\rm d2c} = \dfrac{\dot{\kappa}_{\rm dc}}{x_{\rm s}}
\end{equation}

\begin{figure}
\begin{center}
\includegraphics[width=0.7\linewidth]{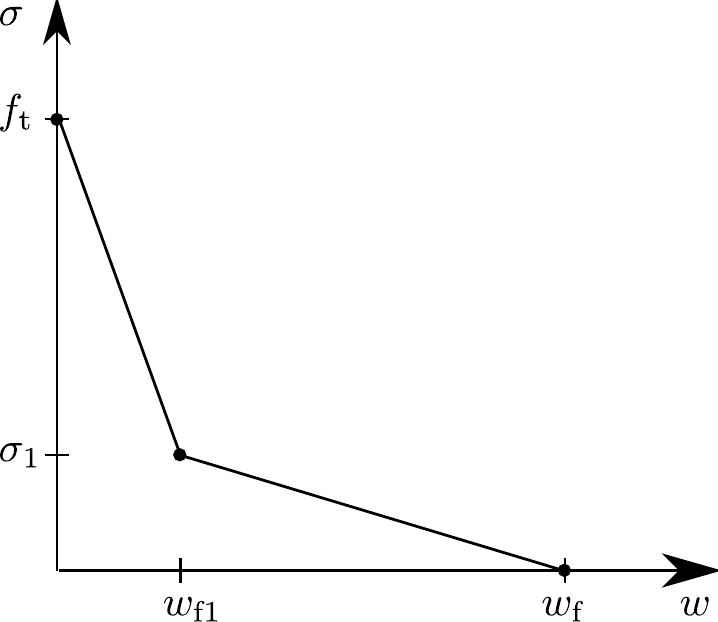}
\end{center}
\caption{Bilinear softening.}
\label{fig:DamBilinear}
\end{figure}

All six damage history variables are used to determine the two damage parameters $\omega_{\rm t}$ and $\omega_{\rm c}$ in Eq.~(\ref{eq:general}).
The form of the two damage parameters $\omega_{\rm t}$ and $\omega_{\rm c}$ depends on the type of softening relation that is modelled.
For instance, for linear softening in uniaxial tension, the stress inelastic displacement relationship in the softening regime is 
\begin{equation}\label{eq:linearCrackOpening}
\sigma = \left \{ \begin{array}{ll} f_{\rm t} \left(1- \dfrac{h \varepsilon_{\rm i}}{w_{\rm f}}\right) & \mbox{if $0 \leq h \varepsilon_{\rm i} < w_{\rm f}$}\\
0 & \mbox{if $ w_{\rm f} \leq  h \varepsilon_{\rm i}$} \end{array} \right .
\end{equation}
where $w_{\rm f}$ is the crack opening at which the uniaxial stress is equal to zero.
To extend this relationship to general loading, the inelastic strain $\varepsilon_{\rm i}$ is replaced by the damage history variables as
\begin{equation}\label{eq:inelasticStrain}
\varepsilon_{\rm i} = \kappa_{\rm d1t}+ \omega_{\rm t} \kappa_{\rm d2t}
\end{equation}
Furthermore, the general stress-strain relationship in Eq.~(\ref{eq:general}) reduces for uniaxial tension to
\begin{equation}\label{eq:linearStressStrain}
\sigma = \left(1-\omega_{\rm t}\right) E \kappa_{\rm dt}
\end{equation}
where $\varepsilon$ is replaced by $\kappa_{\rm dt}$ to make it applicable to general strain states.
Setting Eqs.~(\ref{eq:linearCrackOpening})~and~(\ref{eq:linearStressStrain}) equal gives
\begin{equation}
\omega_{\rm t} = \dfrac{f_{\rm t} \kappa_{\rm d1t} h + w_{\rm f} E \kappa_{\rm dt} - w_{\rm f} f_{\rm t}}{\kappa_{\rm dt} w_{\rm f} E -  f_{\rm t} \kappa_{\rm d2t} h}
\end{equation}
The expression for $\omega_{\rm c}$ is obtained, if $\kappa_{\rm dt}$, $\kappa_{\rm d1t}$ and $\kappa_{\rm d2t}$ are replaced by $\kappa_{\rm dc}$, $\kappa_{\rm d1c}$ and $\kappa_{\rm d2c}$, respectively.
Thus, 
\begin{equation}
\omega_{\rm c} = \dfrac{f_{\rm t} \kappa_{\rm d1c} h + w_{\rm f} E \kappa_{\rm dc} - w_{\rm f} f_{\rm t}}{\kappa_{\rm dc} w_{\rm f} E -  f_{\rm t} \kappa_{\rm d2c} h}
\end{equation}
This procedure can be extended to derive the damage parameter for any combination of line segments, such as the bilinear stress crack opening curve shown in Figure~\ref{fig:DamBilinear}, which is used in the analyses presented in this paper in Sections~2.5~and~3.
For these analyses the parameters of the bilinear stress-strain curve were chosen as $\sigma_1 = 0.2 f_{\rm t}$ and $w_{\rm f1} = 0.2 w_{\rm f}$, for which $w_{\rm f} = 4.444 G_{\rm Ft}/f_{\rm t}$. For all the experiments, the axial specimen length was assumed to be $\ell_{\rm ref} = 0.1$~m for presenting the stress-strain curves.

\subsection{Rate dependence}\label{sec:rate}
The response of concrete is strongly rate dependent. 
If the loading rate is increased, the apparent tensile and compressive strength increase. 
This increase is more pronounced in tension than in compression.
In the present model, this rate dependence is taken into account in Eq.~(\ref{eq:kappadt})~and~(\ref{eq:kappadc}) by the factor $\alpha_{\rm r} \geq 1$.
The greater the rate factor $\alpha_{\rm r}$, the greater is the delay of the onset of damage and, therewith, the strength. 
For tension, the rate dependence is modeled using the expressions proposed by Malvar and Ross in \cite{MalRos98}.
For compression, the expressions reported in CEB-FIP Model Code 1990 \cite{CEB91} are used.
Accordingly, the factor $\alpha_{\rm r}$ is defined as
\begin{equation} \label{eq:rateFactor}
\alpha_{\rm r} = \left(1-\alpha_{\rm c}\right) \alpha_{\rm rt} + \alpha_{\rm c} \alpha_{\rm rc}
\end{equation}
where $\alpha_{\rm c}$ is the factor introduced in Eq.~(\ref{eq:alpha}) to model the different strain rate dependence in tension and compression.

For tension, the rate factor is
\begin{equation}
\alpha_{\rm rt} = \left\{ \begin{array}{ll} 1 & \mbox{ for $\dot{\varepsilon}_{\rm max} \leq 30 \times 10^{-6}$~s$^{-1}$}\\
 \left(\dfrac{\dot{\varepsilon}_{\rm max}}{\dot{\varepsilon}_{\rm t0}}\right)^{\delta_{\rm s}} & \mbox{ for $30 \times 10^{-6}$~s$^{-1} \leq  \dot{\varepsilon}_{\rm max} \leq 1$~s$^{-1}$}\\
 \beta_{\rm s} \left(\dfrac{\dot{\varepsilon}_{\rm max}}{\dot{\varepsilon}_{\rm t0}}\right)^{1/3} & \mbox{ for  $1$~s$^{-1} \leq \dot{\varepsilon}_{\rm max} $}  \end{array} \right.
\end{equation}
with
\begin{equation}
\delta_{\rm s} = \dfrac{1}{1 + 8 f_{\rm c}/f_{\rm c0}}
\end{equation}
and
\begin{equation}
\log{\beta_{\rm s}} = 6 \delta_{\rm s} - 2
\end{equation}
where $\dot{\varepsilon}_{\rm max}$ is the maximum principal strain rate component, $f_{\rm c0} = 10$~MPa and $\dot{\tilde{\varepsilon}}_{\rm t0} = 1 \times 10^{-6}$~s$^{-1}$. For uniaxial tension, for instance, the maximum principal strain rate component $\dot{\varepsilon}_{\rm max}$ is equal to the strain rate in uniaxial tension.
For compression, the rate factor is
\begin{equation}
\alpha_{\rm rc} = \left\{ \begin{array}{ll} 1 & \mbox{ for $\left |\dot{\varepsilon}_{\rm min}\right | \leq 30 \times 10^{-6}$~s$^{-1}$}\\
\left(\dfrac{\|\dot{\varepsilon}_{\rm min}\|}{\dot{\varepsilon}_{\rm c0}}\right)^{1.026\alpha_{\rm s}}  & \mbox{ for $30 \times 10^{-6}$~s$^{-1} \leq  \left |\dot{\varepsilon}_{\rm min}\right | \leq 30$~s$^{-1}$}\\
 \gamma_{\rm s} \left(\dfrac{\|\dot{\varepsilon}_{\rm min}\|}{\dot{\tilde{\varepsilon}}_{\rm c0}}\right)^{1/3} & \mbox{ for 30~s$^{-1}$ $\leq \left|\dot{\varepsilon}_{\rm min}\right|$ }\end{array} \right.
\end{equation}
where the parameter $\alpha_{\rm s}$ is defined as
\begin{equation}
\alpha_{\rm s} = \dfrac{1}{5+9f_{\rm c}/f_{\rm c0}}
\end{equation}
and
\begin{equation}
\log{\gamma_{\rm s}} = 6.156 \alpha_{\rm s} - 2
\end{equation}
Here, $\dot{\varepsilon}_{\rm min}$ is the minimum principal strain rate component and $\dot{\varepsilon}_{\rm c0} = 30 \times 10^{-6}$~s$^{-1}$.
For uniaxial compression, $\dot{\varepsilon}_{\rm min}$ is equal to the axial compressive strain rate.

\subsection{Model response for  varying strain rates and cyclic loading}
\begin{figure}
\begin{center}
\includegraphics[width=0.7\linewidth]{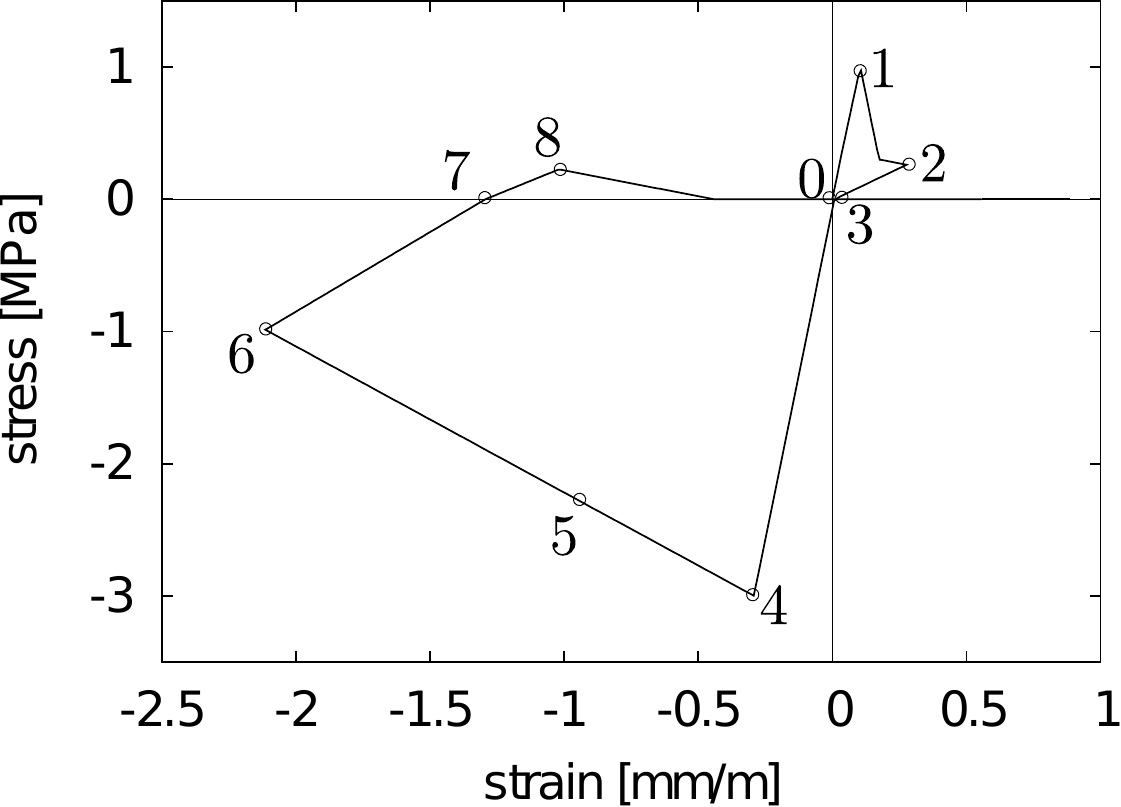}
\end{center}
\caption{Model response for cyclic loading.}
\label{fig:Cyclic}
\end{figure}

The response of the constitutive model is illustrated by several idealised load cases, before it is compared to a wide range of experimental results in the next section.
Firstly, a quasi-static strain cycle is considered, which results in a stress-strain response shown in Fig.~\ref{fig:Cyclic}.
The strain is increased from ``0'' to ``1'', where the tensile strength of the material is reached.
Up to point ``1'', the material response is elastic-plastic with small plastic strains.
With a further increase of the strain form ``1'' to ``2'', the effective stress part continues to increase, since $H_{\rm p}>0$, whereas the nominal stress decreases, since the tensile damage variable $\omega_{\rm t}$ increases.
A reverse of the strain at point ``2'' results in an reduction of the stress with an unloading stiffness, which is less than the elastic stiffness of an elasto-plastic model, but greater than the stiffness of an elasto-damage mechanics model.
At point ``3''when the stress is equal to zero, a further reduction of the strain leads to a compressive response following a linear stress-strain relationship between the points ``3'' and ``4'' with the original Young's modulus $E$ of the undamaged material.
This change of stiffness is obtained by using two damage parameters, $\omega_{\rm t}$ and $\omega_{\rm c}$.
At point ``3'' $\omega_{\rm t}>0$, but $\omega_{\rm c} = 0$.
Up to ``4'' no further plastic strains are generated, since the hardening from ``0'' to ``1'' has increased the elastic domain of the plasticity part so much that the yield surface is not reached at ``4''.
A further decrease of the strain activates the compressive damage variable which leads to a reduction of the nominal stress.
At point ``5'', the plasticity surface is reached and a subsequent increase of strain results in hardening of the plasticity part.
However, the nominal stress, shown in Figure~\ref{fig:Cyclic}, decreases, since $\omega_{\rm c}$ increases.
A second reversal of the strain direction (``6'') changes the stress from tension to compression at ``7'', which is again associated with a change of the stiffness.

\begin{figure}
\begin{center}
\includegraphics[width=0.7\linewidth]{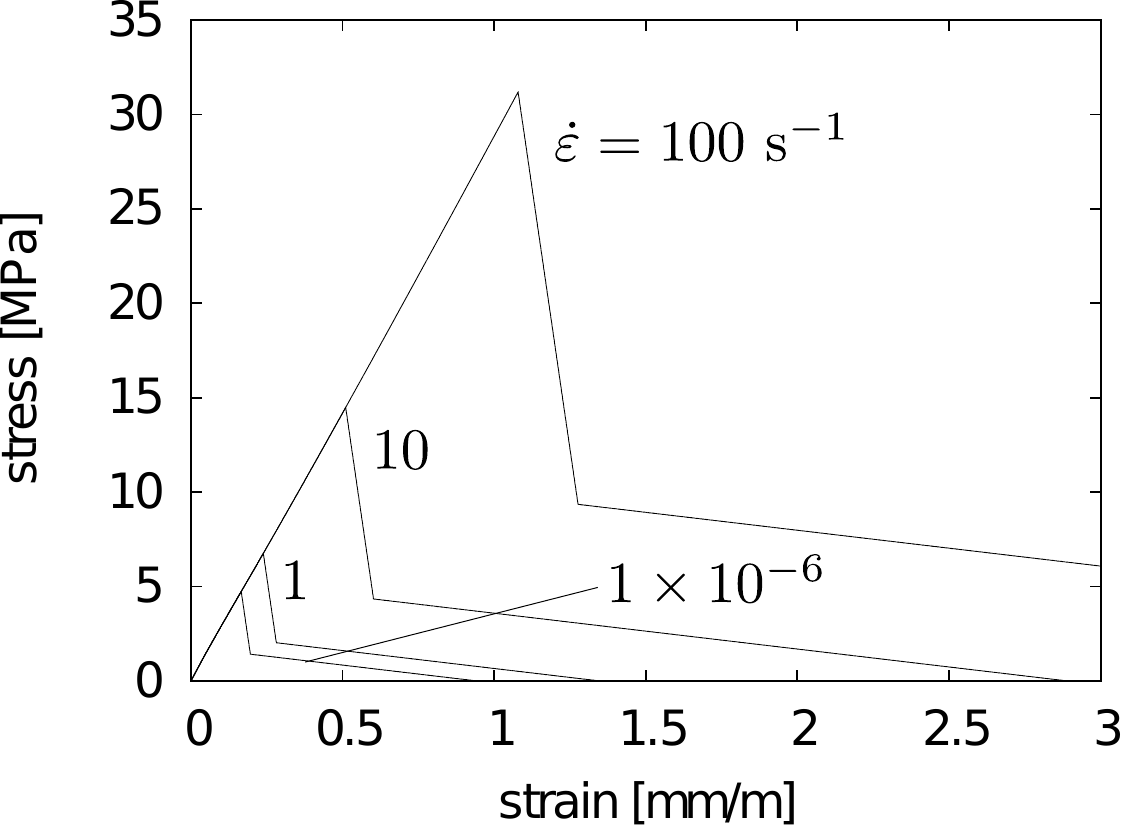}
\end{center}
\caption{Rate effect in tension: Stress strain response for four tensile strain rates.}
\label{fig:RateTension}
\end{figure}

The second group of examples consists of several tensile and compressive loading cases with constant strain strain rates.
For uniaxial tension, strain rates of $1\times 10^{-6}$, $1$, $10$ and $100$~1/s are applied.
The corresponding stress strain responses are shown in Fig.~\ref{fig:RateTension}.
An increase of the loading rate results in a delay of the onset of damage. The strength is increased by factor $\alpha_{\rm r}$ in Eq.~(\ref{eq:rateFactor}), whereas the fracture energy is increased by the factor $\alpha_{\rm r}^2$. The initial stiffness is modelled to be independent of the strain rate.

\begin{figure}
\begin{center}
\includegraphics[width=0.7\linewidth]{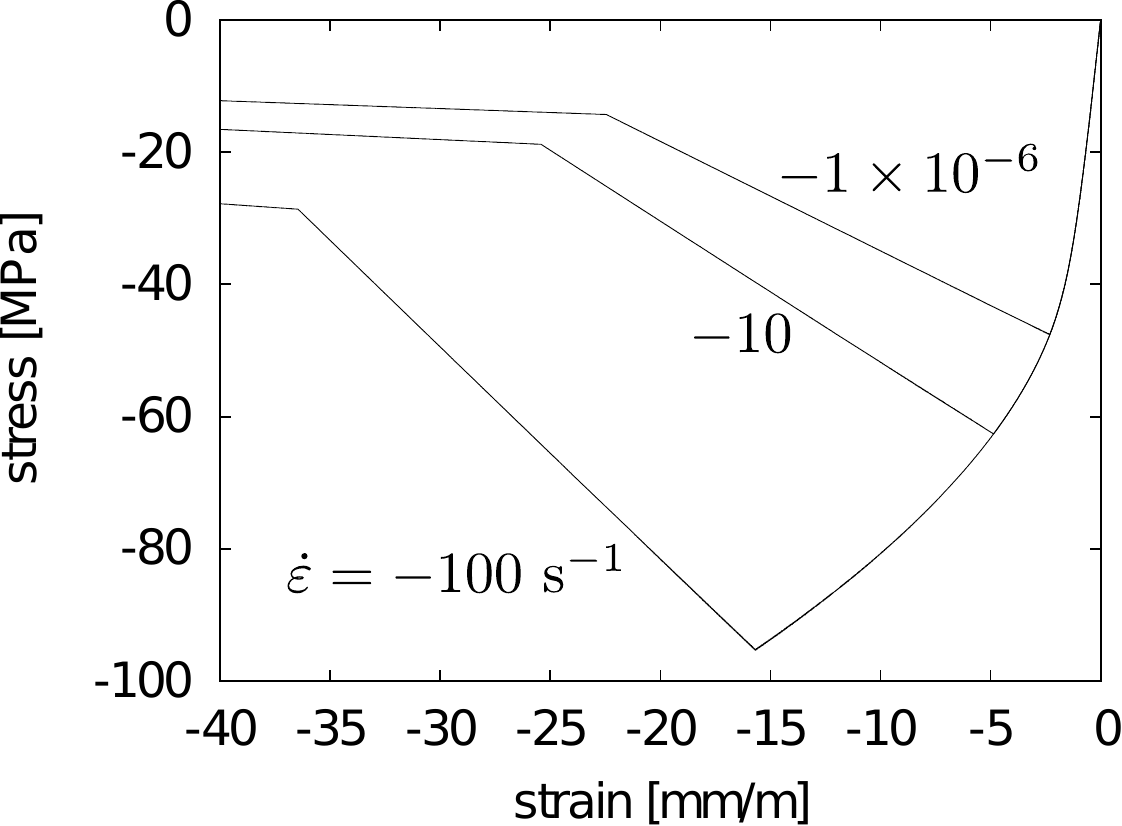}
\end{center}
\caption{Rate effect in compression: Stress strain response for three compressive strain rates.}
\label{fig:RateCompression}
\end{figure}

For uniaxial compression, strain rates of $-1\times 10^{-6}$, $-10$ and $-100$ are considered. The results are shown in Figure~\ref{fig:RateCompression}.
Again, the strength in compression is increased by the factor $\alpha_{\rm r}$ and the fracture energy in compression by $\alpha_{\rm r}^2$.
In compression, the strain at peak stress for greater strain rates than for quasistatic loading depends strongly on the modulus $H_{\rm p}$ of the hardening function $q_{\rm s}$ depicted in Fig.~\ref{fig:hardening}. An increase of the strain rate results in a delay of the onset of damage. This delay shifts the point of peak stress from intersection point of the hardening functions $q_{\rm h}$ and $q_{\rm s}$ into the hardening regime of $q_{\rm s}$.

\section{Comparison with experimental results}

\begin{figure}[htb!]
\begin{center}
\includegraphics[width=0.8\linewidth]{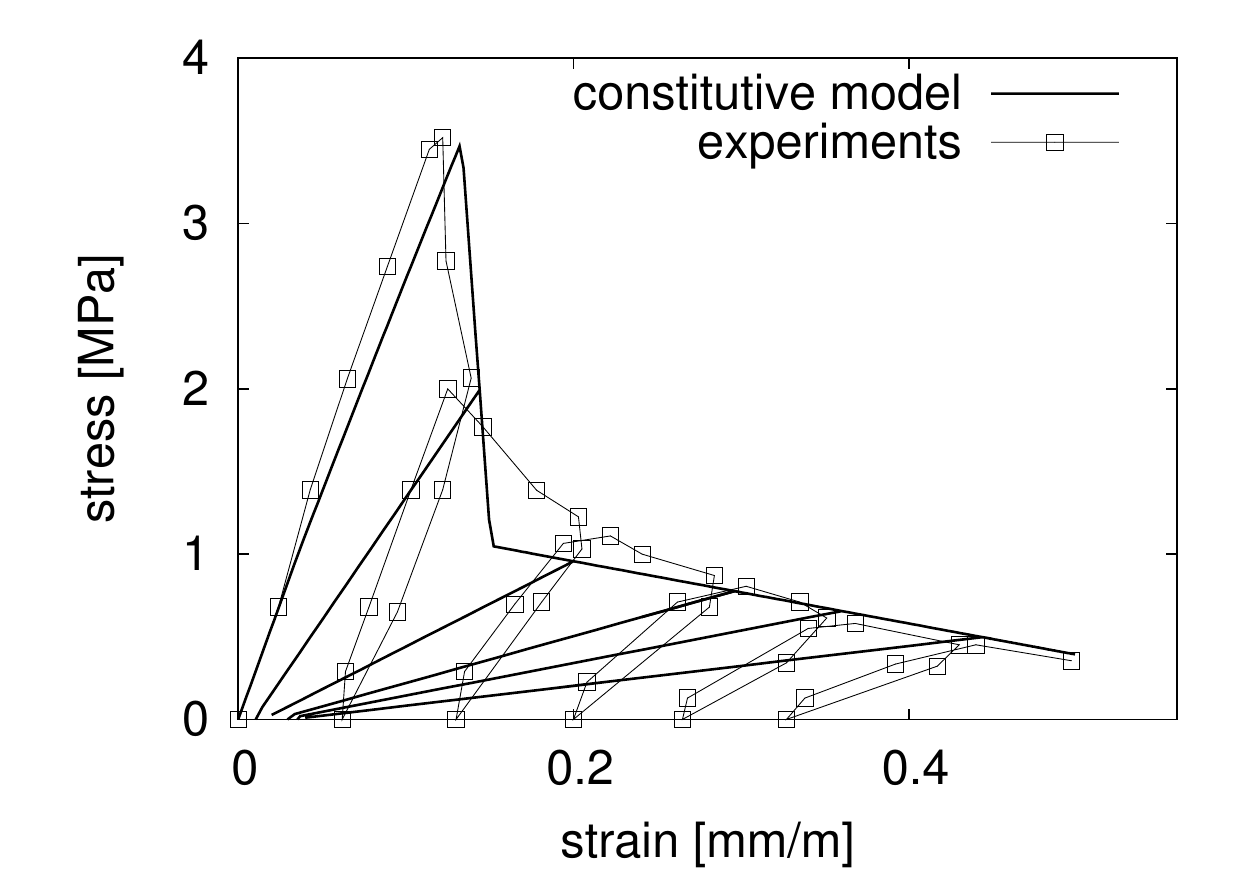}
\end{center}
\caption{Uniaxial tension: Model response compared to experimental results in \protect \cite{GopSha85}.}
\label{fig:GopSha85}
\end{figure}

In this section the model response is compared to five groups of experiments reported in the literature.
For each group of experiments, the model constants $E$, $\nu$, $f_{\rm c}$, $f_{\rm t}$, $G_{\rm Ft}$ and $G_{\rm Fc}$ are adjusted to obtain a fit for the different types of concrete used in the experiments. The other model parameters are set to their default values. The hardening modulus  (shown in Figure~\ref{fig:hardening}) is set to $H_{\rm p} = 0.5$.
The first analysis is uniaxial tensile setup with unloading. The model response is compared to the experimental results reported by Gopalaratnam and Shah in~\cite{GopSha85} (Figure~\ref{fig:GopSha85}).
The relevant model constants for this experiment are $E = 25$~GPa, $\nu = 0.2$, $f_{\rm c} = 40$~MPa, $f_{\rm t} = 3.5$~MPa, $G_{\rm Ft} = 55$~J/m$^2$. 

\begin{figure}[htb!]
\begin{center}
\includegraphics[width=0.8\linewidth]{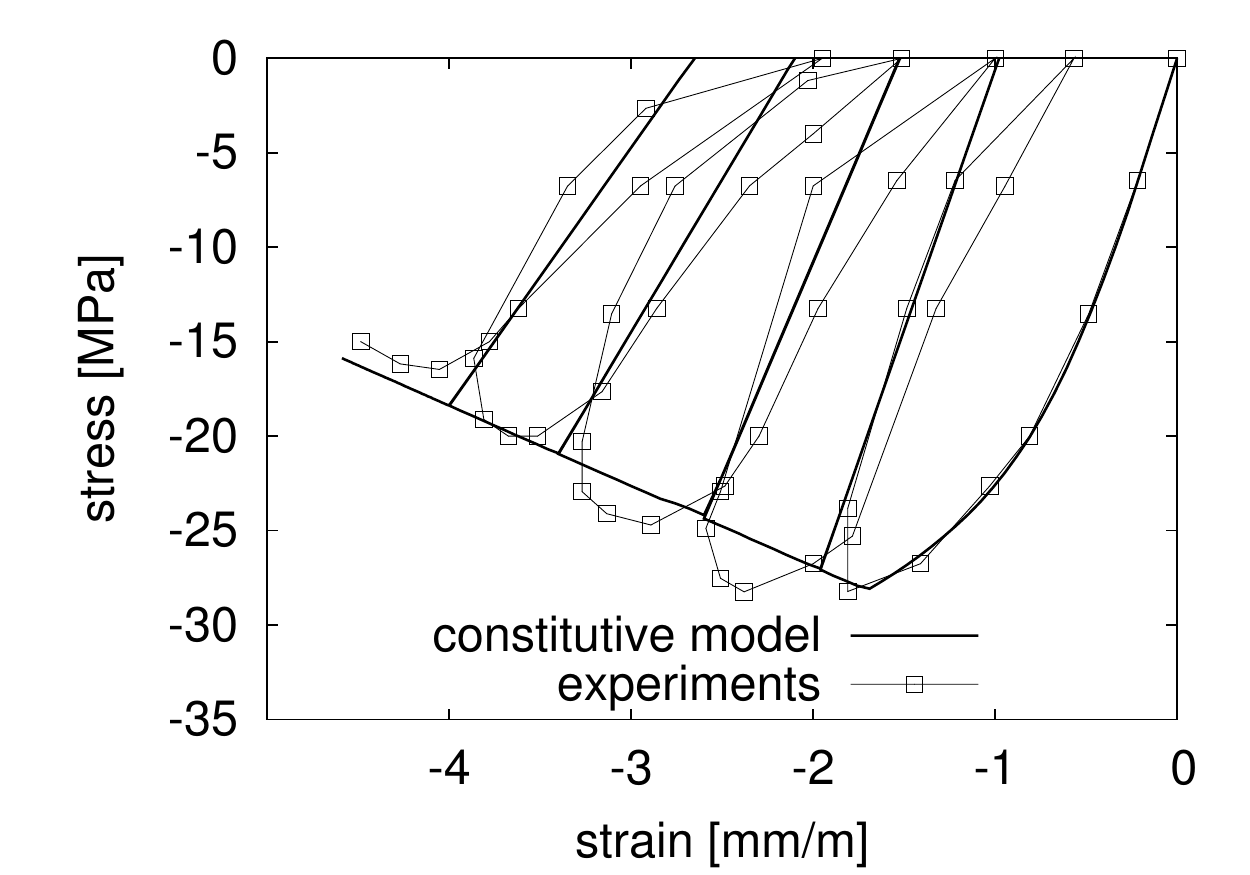}
\end{center}
\caption{Uniaxial compression: Model response compared to experimental results reported in \protect \cite{KarJir69}.}
\label{fig:KarJir69}
\end{figure}

The next example is an uniaxial compression test with unloading, for which the model response is compared to experimental results reported by Karsan and Jirsan \cite{KarJir69} (Figure~\ref{fig:KarJir69}).
The model constants are $E = 30$~GPa, $\nu = 0.2$, $f_{\rm c} = 28$~MPa, $f_{\rm t} = 2.8$~MPa, $G_{\rm Fc} = 2205$~J/m$^2$ .

\begin{figure}[htb!]
\begin{center}
\includegraphics[width=0.8\linewidth]{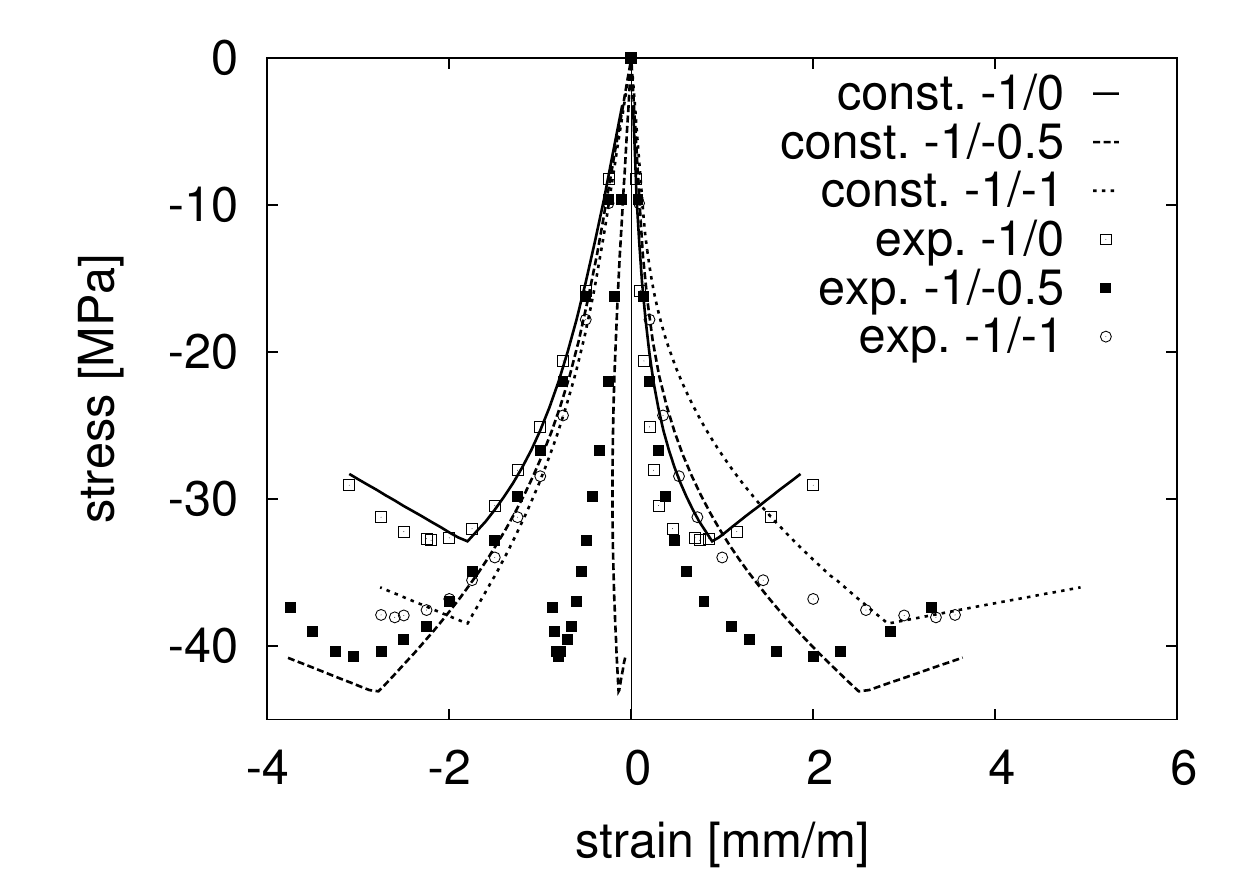}
\end{center}
\caption{Uniaxial compression: Model response compared to experimental results reported in \protect \cite{Kupfer69}.}
\label{fig:Kupfer69}
\end{figure}

Next, the model is compared to uniaxial and biaxial compression tests reported by Kupfer et al. in \cite{Kupfer69}.
For these experiments, the model constants are set to $E = 32$~GPa, $\nu = 0.2$, $f_{\rm c} = 32.8$~MPa, $f_{\rm t} = 3.3$~MPa, $G_{\rm Ft} = 50$~J/m$^2$, $G_{\rm Fc} = 3500$~J/m$^2$. The comparison with experimental results is shown in Figure~\ref{fig:Kupfer69} for uniaxial, equibiaxial and biaxial compression.
For the biaxial compression case, the stress ratio of the two compressive stress components is $\sigma_1/\sigma_2 = -1/-0.5$.

\begin{figure}[htb!]
\begin{center}
\includegraphics[width=0.8\linewidth]{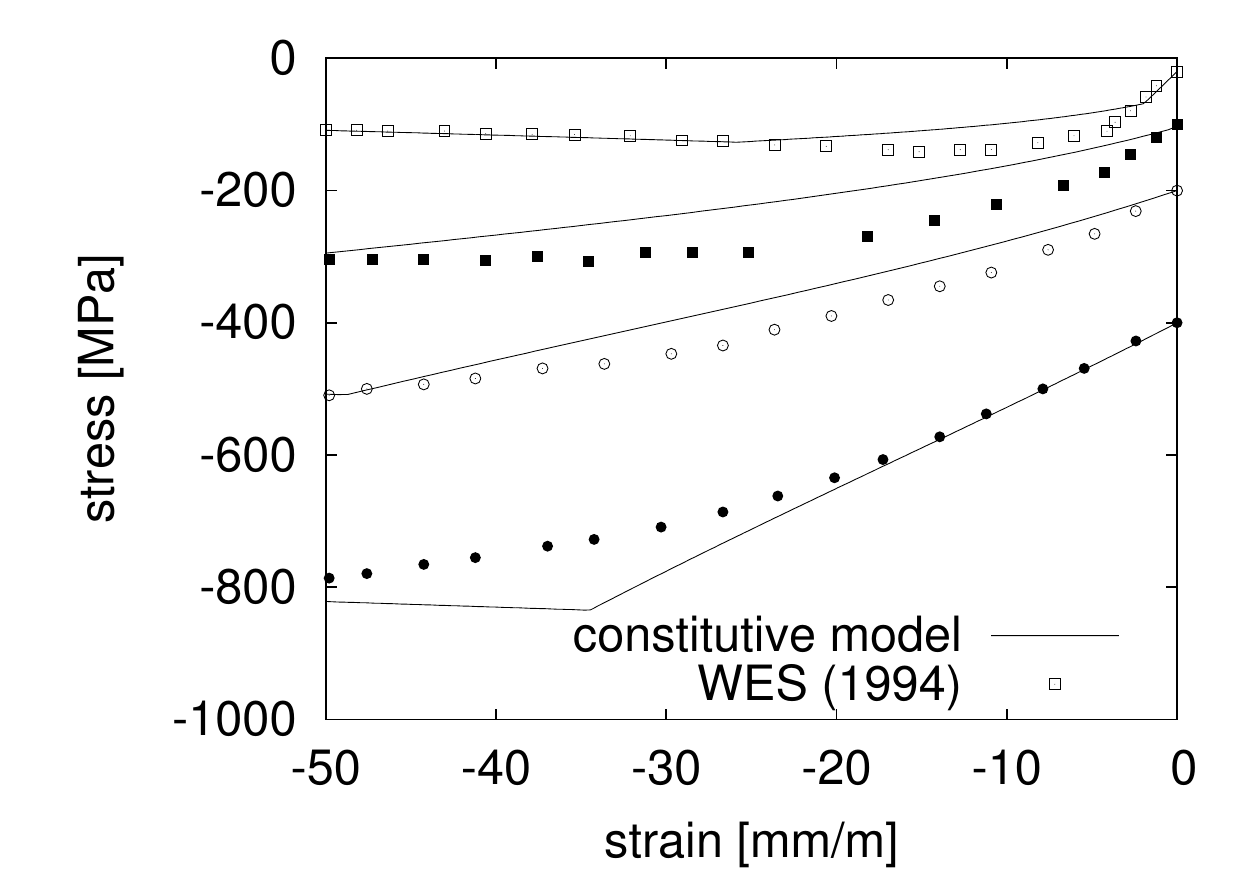}
\end{center}
\caption{Confined compression: Model response compared to experiments used in \protect \cite{CanBaz00}.}
\label{fig:WES1994Tri}
\end{figure}

\begin{figure}[htb!]
\begin{center}
\includegraphics[width=0.8\linewidth]{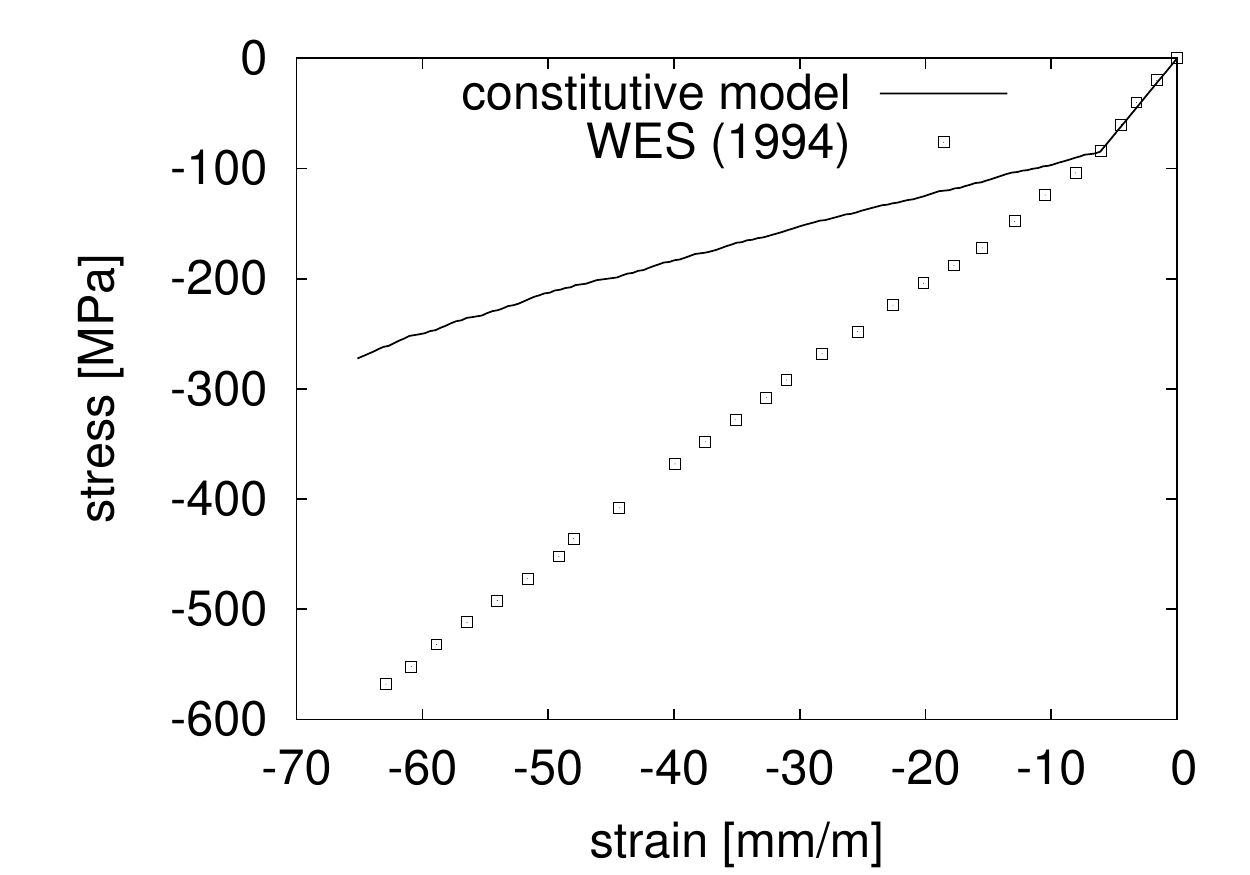}
\end{center}
\caption{hydrostatic compression: Model response compared to experiments used in \protect \cite{CanBaz00}.}
\label{fig:WES1994Hydro}
\end{figure}

Furthermore, the performance of the model is evaluated for triaxial tests and a hydrostatic test reported in \cite{CanBaz00}. 
The material constants for this test are $E = 25$~GPa, $\nu=0.2$,  $f_{\rm c} = 28$~MPa, $f_{\rm t} = 2.8$~MPa, $G_{\rm Fc} = 2205$~J/m$^2$.  
The model response is compared to experimental results presented in Figures~\ref{fig:WES1994Tri}~and~\ref{fig:WES1994Hydro}.

Finally, the model response in triaxial compression is compared to the experimental results reported in \cite{ImrPan96}.
The material constants for this test are $E = 25$~GPa, $\nu=0.2$,  $f_{\rm c} = 28$~MPa, $f_{\rm t} = 2.8$~MPa, $G_{\rm Ft} = 100$~J/m$^2$, $G_{\rm Fc} = 15000$~J/m$^2$.  
\begin{figure}[htb!]
\begin{center}
\includegraphics[width=0.8\linewidth]{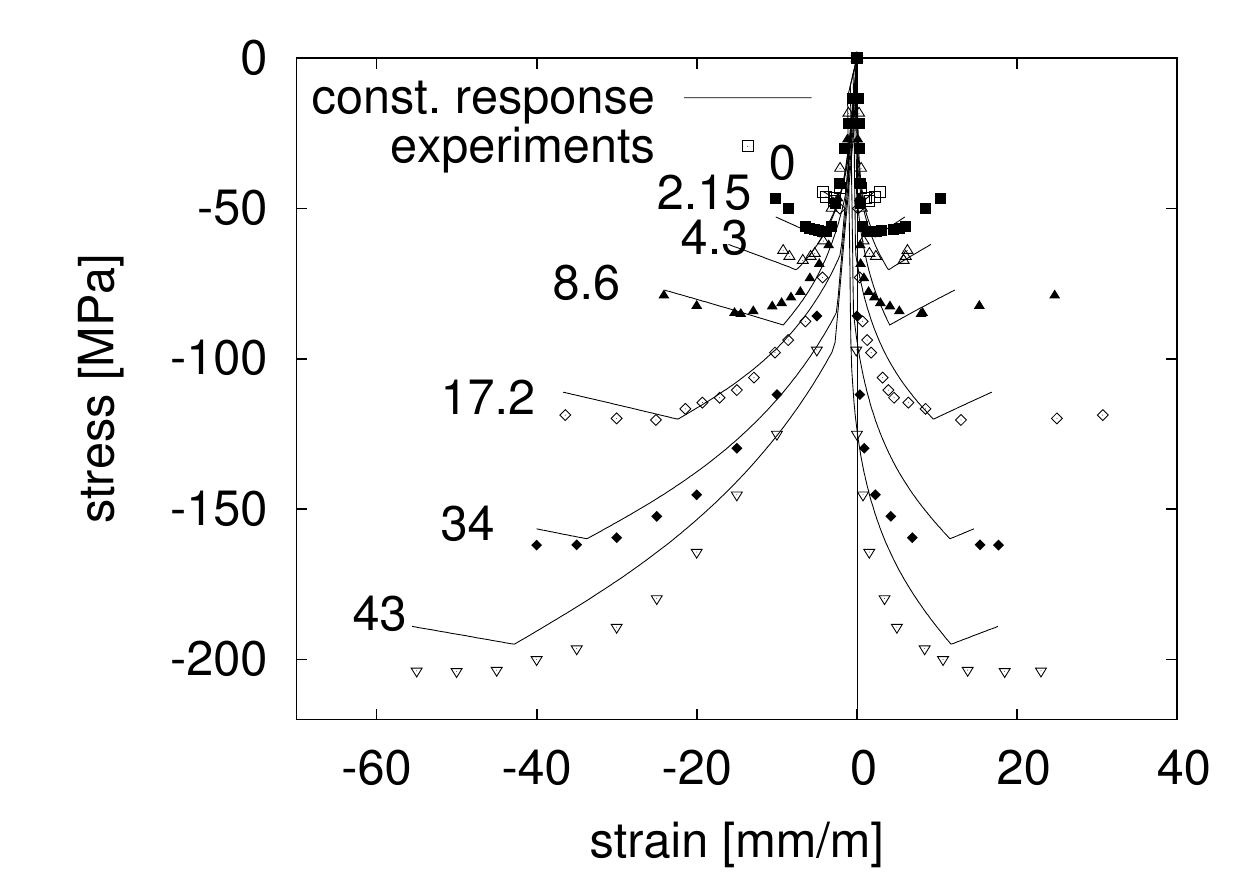}
\end{center}
\caption{Confined compression: Model response compared to experiments reported in \protect \cite{ImrPan96}.}
\label{fig:ImrPan96}
\end{figure}

Overall, the agreement of the model response with the experimental results is very good. 
The model is able to represent the strength of concrete in tension and multiaxial compression. 
In addition, the strains at maximum stress in tension and compression agree well with the experimental results.
The bilinear stress-crack opening curve that was used results in a good approximation of the softening curve in uniaxial tension and compression.

\section{Conclusions}

The present damage plasticity model combines a stress-based plasticity part with a strain based damage mechanics model. 
The model response is in good agreement with experimental results for a wide range of loading from uniaxial tension to confined compression.
In the next steps, the model will be applied to boundary value problems to evaluate its performance to describe failure processes of concrete mesh-independently. It is expected that the model will perform well for these problems since the softening response is formulated so that the crack band approach can be applied, which is known to lead in many cases to mesh-independent results.

\section*{Acknowledgments}
The work is performed within the project "Dynamic behaviour of reinforced concrete structures subjected to blast and fragment impacts" which in turn is financially sponsored by MSB - the Swedish Civil Contingencies Agency.






\bibliographystyle{eurodyn2011}
\bibliography{general}

\begin{thebibliography}{10}

\bibitem{Mazars84}
J.~Mazars,
\newblock ``Application de la m\'{e}canique de l'endommagement au comportement
  non lin\'{e}aire et \`{a} la rupture du b\'{e}ton de structure,''
\newblock {Th\`{e}se de Doctorat d'Etat}, Universit\'{e} Paris VI., France,
  1984.

\bibitem{Ju89}
J.~W. Ju,
\newblock ``On energy-based coupled elastoplastic damage theories:
  {C}onstitutive modeling and computational aspects,''
\newblock {\em \IJSS}, vol. 25, no. 7, pp. 803--833, 1989.

\bibitem{LeeFen98}
J.~Lee and G.~L. Fenves,
\newblock ``Plastic-damage model for cyclic loading of concrete structures,''
\newblock {\em \JEM}, vol. 124, pp. 892--900, 1998.

\bibitem{JasHuePijGha06}
L.~Jason, A.~Huerta, G.~Pijaudier-Cabot, and S.~Ghavamian,
\newblock ``{An elastic plastic damage formulation for concrete: Application to
  elementary tests and comparison with an isotropic damage model},''
\newblock {\em Computer Methods in Applied Mechanics and Engineering}, vol.
  195, no. 52, pp. 7077--7092, 2006.

\bibitem{GraJir06}
P.~Grassl and M.~Jir{\'a}sek,
\newblock ``{Damage-plastic model for concrete failure},''
\newblock {\em International Journal of Solids and Structures}, vol. 43, no.
  22-23, pp. 7166--7196, 2006.

\bibitem{GraRem08}
P.~Grassl and R.~Rempling,
\newblock ``A damage-plasticity interface approach to the meso-scale modelling
  of concrete subjected to cyclic compressive loading,''
\newblock {\em Engineering Fracture Mechanics}, vol. 75, pp. 4804--4818, 2008.

\bibitem{Gra09b}
P.~Grassl,
\newblock ``On a damage-plasticity approach to model concrete failure,''
\newblock {\em Proceedings of the ICE - Engineering and Computational
  Mechanics}, vol. 162, pp. 221--231, 2009.

\bibitem{GraJir06a}
P.~Grassl and M.~Jir\'{a}sek,
\newblock ``A plastic model with nonlocal damage applied to concrete,''
\newblock {\em International Journal for Numerical and Analytical Methods in
  Geomechanics}, vol. 30, pp. 71--90, 2006.

\bibitem{Ort87}
M.~Ortiz,
\newblock ``An analytical study of the localised failure modes of concrete,''
\newblock {\em Mechanics of Materials}, vol. 6, pp. 159--176, 1987.

\bibitem{FicBorPij99}
S.~Fichant, C.~La Borderie, and G.~Pijaudier-Cabot,
\newblock ``Isotropic and anisotropic descriptions of damage in concrete
  structures,''
\newblock {\em Mechanics of Cohesive-Frictional Materials}, vol. 4, pp.
  339--359, 1999.

\bibitem{Willam74}
K.~J. Willam and E.~P. Warnke,
\newblock ``Constitutive model for the triaxial behavior of concrete,''
\newblock in {\em Concrete Structures Subjected to Triaxial Stresses}, vol.~19
  of {\em IABSE Report}, pp. 1--30. International Association of Bridge and
  Structural Engineers, Zurich, May 1974.

\bibitem{JirBaz01}
M.~Jir\'{a}sek and Z.~P. Ba\v{z}ant,
\newblock {\em Inelastic Analysis of Structures},
\newblock John Wiley and Sons, Chichester, 2002.

\bibitem{Menetrey95}
Ph. Men\'{e}trey and K.~J. Willam,
\newblock ``{A triaxial failure criterion for concrete and its
  generalization},''
\newblock {\em \JACIS}, vol. 92, pp. 311--318, 1995.

\bibitem{Gra04}
P.~Grassl,
\newblock ``Modelling of dilation of concrete and its effect in triaxial
  compression,''
\newblock {\em Finite elements in analysis and design}, vol. 40, pp.
  1021--1033, 2004.

\bibitem{MalRos98}
L~J Malvar and C~A Ross,
\newblock ``{Review of strain rate effects for concrete in tension},''
\newblock {\em {ACI Materials Journal}}, vol. {95}, no. {6}, pp. {735--739},
  {NOV-DEC} {1998}.

\bibitem{CEB91}
CEB,
\newblock {\em {CEB-FIP Model Code 1990, Design Code}},
\newblock Thomas Telford, London, 1991.

\bibitem{GopSha85}
V.S. Gopalaratnam and S.P. Shah,
\newblock ``{Softening Response of Plain Concrete in Direct Tension},''
\newblock {\em ACI Journal Proceedings}, vol. 82, no. 3, 1985.

\bibitem{KarJir69}
I.~D. Karsan and J.~O. Jirsa,
\newblock ``Behavior of concrete under compressive loadings,''
\newblock {\em \JSD}, vol. 95, pp. 2543--2563, 1969.

\bibitem{Kupfer69}
H.~Kupfer, H.~K. Hilsdorf, and H.~{R\"{u}sch},
\newblock ``{Behavior of concrete under biaxial stresses},''
\newblock {\em \JACI}, vol. 66, pp. 656--666, 1969.

\bibitem{CanBaz00}
F.~C. Caner and Z.~P. Ba\v{z}ant,
\newblock ``Microplane model {M4} for concrete: {II.~Algorithm} and
  calibration,''
\newblock {\em \JEM}, vol. 126, pp. 954--961, 2000.

\bibitem{ImrPan96}
I.~Imran and S.~J. Pantazopoulou,
\newblock ``Experimental study of plain concrete under triaxial stress,''
\newblock {\em \JACIM}, vol. 93, pp. 589--601, 1996.

\end{thebibliography}

\end{document}